\documentclass[
reprint,
showkeys,
amsmath,amssymb,
aps,
prb,
]{revtex4-2}

\usepackage{xcolor}
\usepackage{graphicx}
\usepackage{bm}

\begin{document}
	
\title{Computational Frameworks for Patterned Two-Dimensional Magnetism}
	
\author{Soham Chandra}
\affiliation{Department of Physics, Brainware University, Kolkata 700125, INDIA}
\author{Soumyajit Sarkar}
\email{dss.ph@brainwareuniversity.ac.in}
\affiliation{Department of Physics, Brainware University, Kolkata 700125, INDIA}

\date{\today}
	
\begin{abstract}
Patterned two-dimensional (2D) magnetic nanostructures constitute geometry-engineered spin systems in which exchange, anisotropy, 
dipolar coupling, and finite-size effects operate on comparable energy scales. Spatial modulation of continuous magnetic films produces 
confinement-driven critical behavior, compensation phenomena, metastable switching pathways, and topologically nontrivial textures such 
as vortices and skyrmions. Computational modeling plays a central role in resolving this complexity, enabling quantitative construction 
of thermodynamic phase diagrams and analysis of geometry-dependent stability regimes. This review synthesizes theoretical and numerical frameworks for patterned 2D magnetism, including classical spin models, stochastic 
spin dynamics, rare-event methods, and multiscale parameterization informed by first-principles calculations. Representative 
systems-nanodot and antidot arrays, artificial spin-ice lattices, exchange-modulated heterostructures, and patterned van der Waals magnets-
illustrate how geometry functions as an effective thermodynamic control parameter.
Emerging directions in nonequilibrium modeling, multiphysics coupling, and scalable data-centric workflows are discussed in the context 
of predictive phase mapping. Patterned 2D magnetism thus exemplifies geometry as
an effective thermodynamic variable in materials engineering and 
computational statistical physics, with phase stability and controlled spin textures at the core of next-generation spintronic architectures.
\end{abstract}

\keywords{Patterned two-dimensional magnetism, Magnetic nanostructures, Monte Carlo methods, Spin dynamics, Phase diagrams,
Finite-size scaling, Multiscale modeling, Geometry-controlled magnetic order}
\maketitle

\section{Introduction}
\label{sec:Intro}

Two-dimensional (2D) patterned magnetic nanostructures have emerged as a versatile platform for 
investigating geometry-controlled spin phenomena in reduced dimensions. In this context, ``patterned" 
refers to deliberate spatial modulation of a magnetic layer—achieved through lithographic structuring, 
self-assembly, moir\'e engineering, 
compositional variation, or controlled edge definition—resulting in 
periodic or aperiodic confinement of magnetic regions within an otherwise continuous sheet. In contrast to uniform thin films, patterned 2D magnets exhibit spatially nonuniform exchange coupling, anisotropy, and dipolar fields. These features produce magnetic states that are strongly influenced by confinement and broken translational symmetry.

Advances in fabrication techniques have enabled controlled patterning of van der Waals ferromagnets such as Fe$_3$GeTe$_2$ and chromium-based 
halides, where nanoscale structuring modifies coercivity, switching pathways, and domain stability~\cite{Wang2022:ACSNano}. Experimental observations indicate that confinement can induce quantized magnetization reversal and enhanced domain-wall pinning. These phenomena deviate from predictions based solely on continuum micromagnetics. The introduction of geometric inhomogeneity creates localized energy landscapes capable of stabilizing nontrivial 
spin configurations, including skyrmions~\cite{Vousden2016:APL,Muller2017:PRL,Reichhardt2022:RMP,Junquera2023:RMP}, 
merons~\cite{Lin2015:PRB,Vijayakumar2020:PRApplied}, and magnetic bubble states~\cite{Zhang2020:PRResearch}. In patterned geometries, the competition 
among exchange, anisotropy, dipolar interactions, and boundary effects can generate metastable states and reentrant phase behavior absent in extended films.

Recent experiments on patterned Fe$_3$GeTe$_2$ have demonstrated geometry-controlled coercivity and domain evolution, with reversal 
processes governed by edge-mediated nucleation rather than homogeneous rotation~\cite{Tan2018:NatComm}. 
Comparable behavior in nanopatterned Cr$_2$Ge$_2$Te$_6$ and CrI$_3$ confirms that engineered boundaries significantly influence magnetization dynamics~\cite{Powalla2023PRB}. 
These findings point to a general mechanism, that is, in patterned systems, geometry acts as an additional control variable for modulating the balance between domain-wall energy and shape anisotropy. 
Computational modeling has been instrumental in clarifying the microscopic origin of these effects. Atomistic Monte Carlo simulations and spin-dynamics approaches show that periodic modulation of exchange interactions can produce magnetization plateaus, compensation points, and reentrant transitions. These effects are reminiscent of frustrated spin networks~\cite{Vasiliev2018QuanMat,Elidrysy2024EPJB,Liang2025PRB}. 
Such results provide a quantitative 
connection between experimentally observed hysteresis behavior and underlying spin correlations~\cite{Zhou2023NatCommun}.

Patterning also modifies nonequilibrium spin dynamics. Confinement influences domain-wall mobility, skyrmion nucleation, and magnon 
propagation. Experimental studies of patterned nanoarrays report quantized creation and annihilation of skyrmions under moderate 
magnetic fields~\cite{Gilbert2015NatCommun,Du2015NatCommun}. 
Micromagnetic and atomistic simulations indicate that boundary-induced gradients in effective 
anisotropy and Dzyaloshinskii--Moriya interaction (DMI) alter switching pathways and affect thermal stability~\cite{Varentsova2018,Koraltan2023PRB}. 
Hybrid heterostructures 
further expand the design space: proximity coupling in Fe$_3$GeTe$_2$/graphene and CrI$_3$/hBN bilayers enables electric-field and 
strain-mediated control of magnetization~\cite{Deng2018Nature,Jiang2018NatNanotech,Cheng2019Science}. 
Computational studies of core--shell patterned geometries reveal compensation 
behavior and artificial ferrimagnetism confined within single atomic layers~\cite{Keller2018SciRep,Lalwani2022ApplEngMat}, 
illustrating how interfacial engineering reshapes phase structure.

Despite rapid progress, patterned 2D magnets present significant theoretical and computational 
challenges. Finite-size effects, long-range dipolar interactions, and sensitivity to edge roughness 
complicate both experimental interpretation and modeling~\cite{Cheng2019Science, Gibertini2019:Nature}. 
Finite-size scaling analyses indicate that 
the Curie temperature $\left(T_c\right)$ may depend non-monotonically on pattern periodicity and anisotropy 
strength~\cite{Courtes2024JMMM}. Reduced coordination at edges can suppress long-range order, whereas 
anisotropy enhancement may partially offset this reduction~\cite{Varentsova2018}. 
Reliable modeling, therefore, requires 
statistical sampling strategies capable of resolving both thermodynamic fluctuations and geometric heterogeneity.\\

Over the past decade, computational approaches have evolved from qualitative simulations toward 
quantitatively informed multiscale workflows. Lattice-based Monte Carlo techniques, stochastic 
Landau--Lifshitz--Gilbert dynamics, and micromagnetic solvers are increasingly combined with first-principles 
parameterization~\cite{Kabiraj2022:Patterns}. GPU-accelerated implementations and data-driven surrogate models trained on 
simulation outputs have enabled more efficient exploration of multidimensional phase 
diagrams~\cite{Hu2025PRM, Hait2025, Guan2025, Moreels2026Nature}. 
These methods facilitate systematic analysis of how geometry, anisotropy, 
and exchange interactions cooperate to stabilize targeted magnetic states.\\

\begin{figure}
    \centering
    \includegraphics[width=0.9\linewidth]{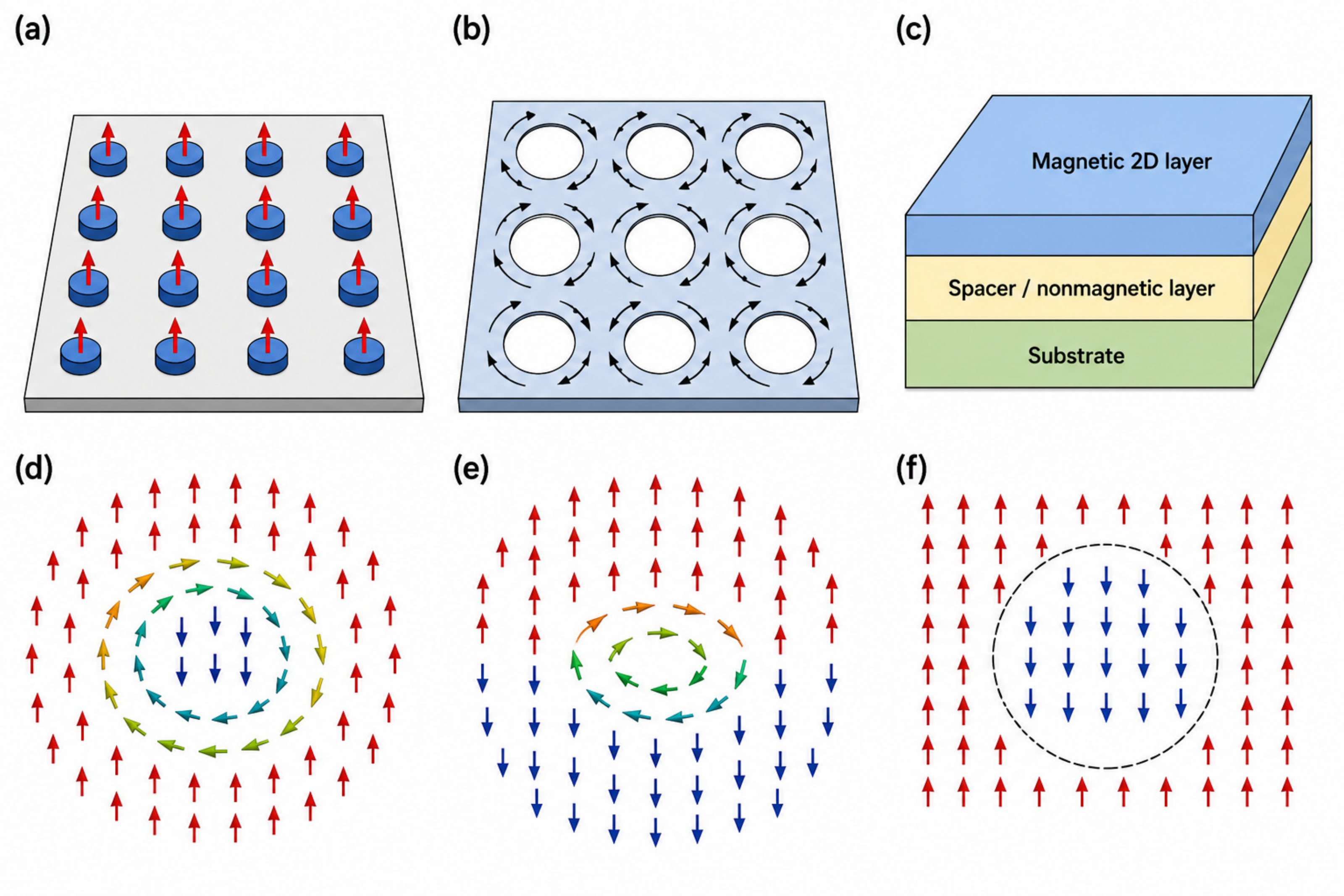}
    \caption{Schematic representations of representative patterned magnetic systems in two-dimensional materials. The upper row illustrates geometrically engineered structures: (a) nanodot arrays, (b) antidot lattices, and (c) hybrid heterostructures. The lower row shows representative magnetic textures: (d) skyrmions, (e) merons, and (f) magnetic bubble domains. The figure is intended to provide a conceptual overview of the principal classes of systems discussed in this review.}
    \label{fig:mag_sys}
\end{figure}

This review presents a computational overview of patterned two-dimensional magnetism, with emphasis on 
methodological developments and their application to thermodynamics, phase behavior, and magnetization 
dynamics. We focus on how confinement, interface engineering, and multiscale coupling modify spin 
order in patterned lattices, and we discuss the reliability and limitations of existing simulation 
frameworks. By organizing recent advances within a unified computational perspective, we aim to 
clarify the principles governing geometry-mediated magnetism in two dimensions and to outline open 
challenges for predictive modeling of patterned spintronic systems. The computational observables discussed in this review map naturally onto experimentally accessible quantities, thereby strengthening the connection between predictive modeling and measurement.
Fig.~\ref{fig:mag_sys} provides schematic representations of several representative systems, including nanodot arrays, antidot lattices, hybrid heterostructures, skyrmions, merons, and magnetic bubble domains, which are discussed throughout this review.

\section{Fundamentals of Patterned Magnetism}
\label{sec:funda_pat_mag}

Patterned magnetism in 2D systems departs fundamentally from bulk behavior -- geometry, topology, and boundary conditions become intrinsic components of the magnetic Hamiltonian~\cite{Wang2023JAP,Cheng2019Science,Burch2020Nature}. In patterned structures—such as nanodot arrays, antidot lattices, nanowires, or moiré-engineered superlattices—a continuous magnetic film is segmented into spatially confined regions whose interactions are mediated by exchange bridges, dipolar coupling, and interface effects. The resulting magnetic state is governed not only by intrinsic material parameters but also by geometric constraints and connectivity.

Finite-size scaling and atomistic simulations demonstrate that confinement modifies both critical behavior and reversal mechanisms~\cite{Dakir2015PhysA,Nina2024PRB}. Reduced coordination at edges, spatial variation of anisotropy, and long-range dipolar interactions introduce effective inhomogeneities that alter correlation lengths and broaden transition signatures. In patterned 2D magnets, geometry therefore acts as an additional control parameter alongside exchange and anisotropy.

\subsection{Exchange, Anisotropy, and Dipolar Interactions}
The energetics of patterned magnets arise from competing exchange, anisotropy, dipolar interactions, external magnetic fields, and, 
where relevant, the Dzyaloshinskii--Moriya interaction (DMI). Spatial modulation of these interactions reflects the inhomogeneity 
introduced by confinement, interfaces, defects, and compositional variations. A generic spin Hamiltonian may be written as
\begin{equation}
\begin{aligned}
\mathcal{H} =\,
&- \sum_{\langle i,j \rangle}
J_{ij}\,\mathbf{S}_i \cdot \mathbf{S}_j
- \sum_i K_i (S_i^z)^2 \\
&- \sum_{\langle i,j \rangle}
\mathbf{D}_{ij}\cdot
(\mathbf{S}_i \times \mathbf{S}_j)
- \sum_i \mu_i\,\mathbf{H}\cdot\mathbf{S}_i \\
&+ \mathcal{H}_{\mathrm{dip}} .
\end{aligned}
\label{eq:Hamiltonian}
\end{equation}
where $\mathbf{S}_i$ denotes the spin vector at lattice site $i$, and $\mathbf{H}$ represents an externally applied magnetic field. 
The first, second, third, fourth, and fifth terms correspond to the exchange, magnetocrystalline anisotropy, Dzyaloshinskii--Moriya, 
Zeeman, and dipole--dipole interactions, respectively. In patterned magnetic systems, the exchange coupling $J_{ij}$, anisotropy 
constant $K_i$, Dzyaloshinskii--Moriya vector $\mathbf{D}_{ij}$, and local magnetic moment $\mu_i$ may vary spatially owing to edge effects, interface asymmetry, strain, or compositional modulation.
Exchange interactions promote local alignment, while anisotropy stabilizes long-range order in two dimensions by opening a 
magnon gap. In ultrathin films, perpendicular magnetic anisotropy often emerges from interface-induced spin--orbit 
coupling~\cite{Tong2025PRApplied}. Patterning redistributes anisotropy locally, particularly at edges and corners. Dipolar 
interactions, intrinsically long range, couple distant regions and strongly depend on element geometry~\cite{Prozorov2018PRApplied}. 
Numerical studies show that shape—circular, square, or triangular—modifies effective demagnetization factors and switching fields. 
When present, DMI introduces chirality and can be spatially engineered through interfacial design~\cite{Zhu2022CommPhys}.

\subsection{Emergent Magnetic Textures}
The competition among exchange, anisotropy, dipolar fields, and confinement stabilizes diverse spin textures in patterned geometries. Nanodisks commonly support vortex states characterized by flux-closure configurations that minimize magnetostatic energy~\cite{Hoffmann2002JAP}. Skyrmionic and bubble-like textures can emerge when chiral interactions or boundary constraints balance exchange stiffness and anisotropy~\cite{Yang2021SciRep}. Artificial spin-ice arrays extend this concept by arranging nanoislands into frustrated geometries that mimic kagome or square ice lattices~\cite{Wang2006Nature,Ortiz-Ambriz2019RMP}. Monte Carlo and micromagnetic simulations reveal emergent monopole-like excitations and long-range dipolar correlations in such systems~\cite{Yumnam2021PRB}. Modeling these textures requires accurate treatment of long-range interactions and disorder, which significantly increase computational complexity.

\subsection{Finite-Size and Boundary Effects}

In nanoscale patterned magnets, the thermodynamic limit is not attainable; finite geometry directly shapes magnetic response. Correlation lengths saturate at the characteristic pattern size $L$, producing rounded magnetization curves and broadened susceptibility peaks near criticality. Finite-size scaling (FSS) provides a quantitative framework for analyzing this dependence, though the effective system size must reflect geometric connectivity rather than simple spin count~\cite{Mordret2024PRB}. Binder cumulant analysis remains a standard method for identifying pseudo-critical temperatures, but its interpretation in disconnected or weakly coupled arrays requires caution~\cite{LandauBinder2014:CUP}.

Edges introduce reduced coordination and modified anisotropy, frequently serving as nucleation sites 
for domain walls or vortices under applied fields. Micromagnetic simulations show that switching 
sequences depend sensitively on pattern symmetry and shape anisotropy~\cite{Fidler2001:IEEE}. In 
connected networks, exchange bridges between neighboring elements create collective reversal modes 
distinct from isolated dots.

Interfaces in multilayer or core--shell geometries further enrich magnetic behavior. Exchange bias, 
interfacial DMI, and charge transfer modify local coupling constants and anisotropy 
energies~\cite{Evans2014JPCM}. First-principles-informed spin models increasingly incorporate such 
interfacial variations to improve quantitative agreement with experiment.

\subsection{Electronic and Structural Coupling}

Magnetism in 2D materials is closely linked to electronic structure. Nanoscale patterning alters 
orbital hybridization, charge redistribution, and spin--orbit coupling, thereby modifying exchange 
parameters and magnetocrystalline anisotropy. Density functional theory (DFT) studies demonstrate that 
lateral confinement and edge termination can significantly enhance anisotropy energies and reshape 
band structures in systems such as CrI$_3$ and Fe$_3$GeTe$_2$~\cite{Gibertini2019:Nature}. 
Edge-localized states and symmetry breaking frequently increase spin polarization and stabilize 
out-of-plane magnetization.

Multiscale modeling frameworks now integrate DFT-derived exchange and anisotropy constants into 
atomistic spin and micromagnetic simulations. This parameter transfer enables consistent treatment of 
electronic, thermodynamic, and dynamic contributions to patterned magnetism~\cite{Evans2014JPCM}. 
Machine-learning-assisted models further accelerate exploration of the high-dimensional design space defined by 
geometry, composition, and interfacial structure~\cite{Merker2022iScience,Xu2025PNAS}.

\subsection{Thermodynamic and Statistical Perspectives}

Thermal fluctuations play a crucial role in two dimensions. Long-range order is stabilized through 
anisotropy and spin--orbit coupling, which introduce an energy gap in the magnon spectrum and 
circumvent the restrictions of the Mermin--Wagner theorem\cite{Mermin1966PRL,Hohenberg1967PhysRev}. 
In patterned systems, geometry modifies 
entropy and correlation length, leading to broadened transitions and geometry-mediated Curie temperatures\cite{Dupont2021PRL}.

Spatially varying exchange and anisotropy can produce compensation points and reentrant phase behavior 
in core--shell or heterostructured arrays. Monte Carlo simulations reveal how sublattice imbalance or 
competing exchange paths generate nonmonotonic temperature dependence of magnetization~\cite{Nmaila2023, Wang2025ApplPhysA}. Canonical 
thermodynamic observables—magnetization, internal energy, specific heat, and susceptibility—remain the 
primary tools for mapping phase boundaries, though interpretation must account for confinement and disorder.
Overall, the fundamentals of patterned magnetism demonstrate that geometry, interface structure, and 
electronic coupling enter on equal footing with intrinsic magnetic interactions. These considerations 
form the physical basis for the computational methodologies discussed in the following section.

\section{Computational Frameworks for Patterned 2D Magnetism}
\label{sec:computational_frameworks}

\begin{figure*}
    \centering
    \includegraphics[width=0.8\linewidth]{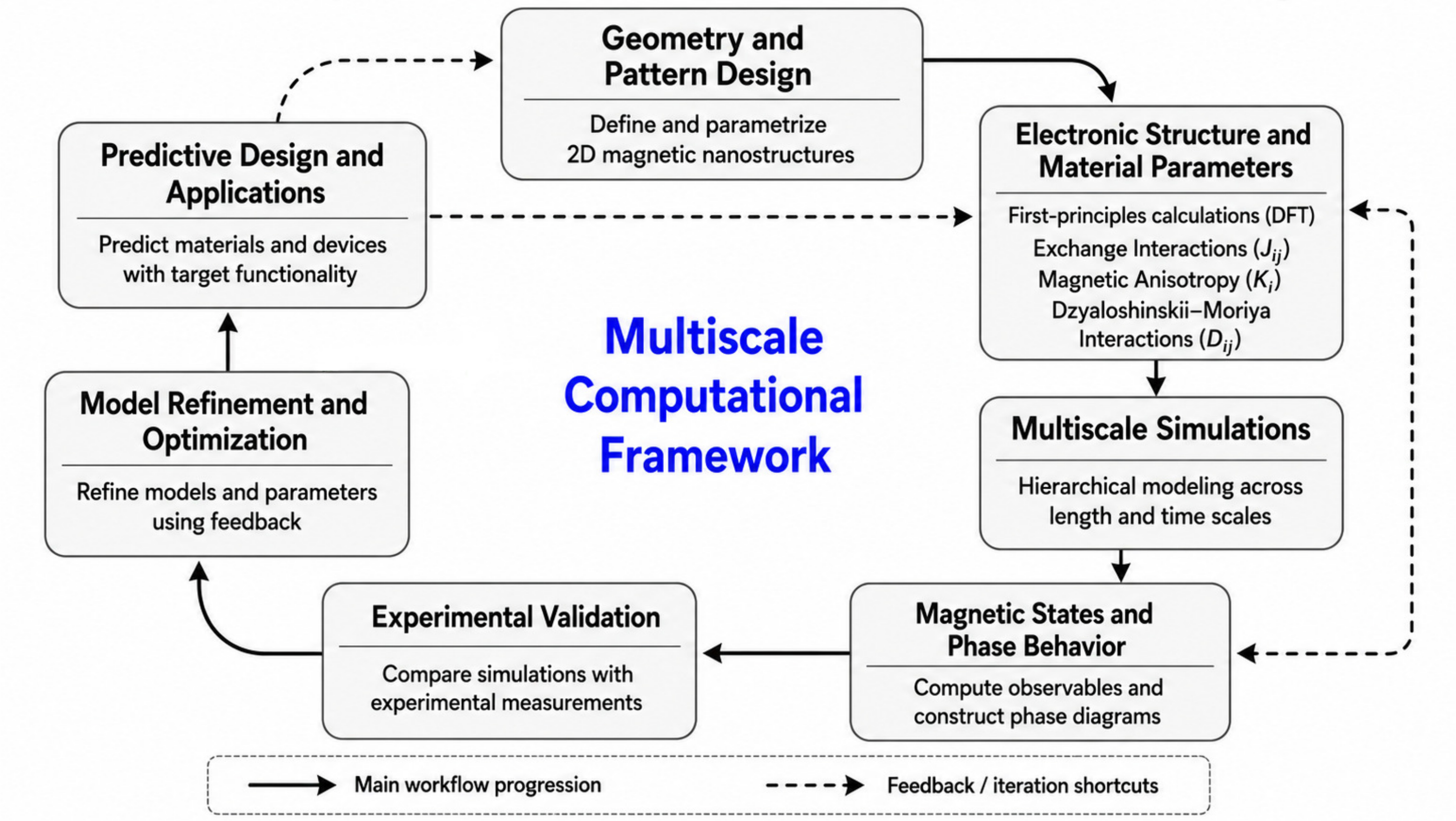}
    \caption{Schematic illustration of a multiscale computational framework for patterned two-dimensional magnetism, 
    linking geometry and pattern design, first-principles electronic-structure calculations, spin-based simulations, 
    magnetic-state characterization, experimental validation, and model refinement. Solid arrows indicate the primary workflow, whereas dashed arrows denote feedback pathways and iterative optimization cycles that facilitate predictive materials and device design.}
    \label{fig:workflow}
\end{figure*}

Patterned 2D magnetic systems present a multiscale problem in which geometry, exchange, anisotropy, dipolar interactions, and 
interface effects must be treated consistently. Computational modeling therefore plays a central role in resolving both 
equilibrium thermodynamics and nonequilibrium magnetization dynamics. The principal numerical approaches include Monte Carlo 
(MC) sampling for equilibrium properties, spin-dynamics simulations based on the Landau--Lifshitz--Gilbert (LLG) equation for 
time-dependent behavior, and multiscale parameterization schemes linking first-principles calculations to atomistic and 
micromagnetic models. Recent progress in high-performance computing and data-driven acceleration has significantly expanded 
accessible length and time scales. The computational investigation of patterned two-dimensional magnetism has consequently 
evolved into a hierarchical and iterative workflow that integrates electronic-structure methods, spin-based simulations, 
experimental characterization, and model refinement.
Fig.~\ref{fig:workflow} provides a schematic overview of this multiscale framework, which 
serves as the foundation for the methodologies discussed in the following sections. Notably, the workflow is not strictly 
linear; information obtained from magnetic-state characterization and experimental measurements can be fed back into the 
modeling pipeline, enabling iterative refinement of computational parameters and improved predictive accuracy.

\subsection{Monte Carlo Methods}

Monte Carlo techniques remain the primary tool for investigating thermodynamic properties of lattice spin models. In the Metropolis algorithm~\cite{Metropolis1953}, trial 
spin updates are accepted with probability $\min\{1,\exp(-\Delta E/k_BT)\}$, ensuring detailed balance and convergence to the Boltzmann distribution. 
Near criticality, critical slowing down reduces efficiency; cluster algorithms such as 
Wolff or Swendsen--Wang alleviate this by updating correlated spin domains simultaneously~\cite{Wolff1989:PRL}.

For patterned magnets, spatial heterogeneity is incorporated through position-dependent interaction parameters and lattice 
geometries that reflect the underlying pattern topology. Monte Carlo simulations are therefore typically performed using the 
generalized spin Hamiltonian of Eq.~(\ref{eq:Hamiltonian}), 
with spatially varying exchange, anisotropy, Dzyaloshinskii--Moriya, dipolar, and 
Zeeman contributions where appropriate. The choice of boundary conditions (periodic, open, or mixed) further enables the modeling of 
confinement effects, interfaces, and finite-size phenomena relevant to patterned magnetic systems.
Long-range dipolar interactions substantially increase the computational cost. Efficient evaluation relies on Ewald summation, 
fast Fourier transform (FFT) techniques, or hierarchical methods 
depending on system size and boundary conditions. 
FSS~\cite{Binder1981} analysis is commonly employed 
to extract effective transition temperatures and critical exponents. The Binder cumulant,
\begin{equation}
U_L = 1 - \frac{\langle M^4\rangle}{3\langle M^2\rangle^2},
\end{equation}
remains a robust indicator of pseudo-critical behavior, though interpretation in spatially inhomogeneous or weakly connected arrays requires careful definition of effective system size.

Beyond equilibrium sampling, generalized ensembles such as Wang--Landau~\cite{Wang2001:PRL} or 
parallel tempering improve exploration of rugged energy landscapes, 
enabling accurate estimation of density of states, compensation points, and reentrant phase behavior in patterned geometries.

\subsection{Spin Dynamics and Magnetization Reversal}

Time-dependent processes in patterned magnets are described using atomistic or micromagnetic spin-dynamics simulations. These simulations are based on the Landau--Lifshitz--Gilbert equation~\cite{Lakshmanan2011:PTRSA},
\begin{equation}
\frac{d\mathbf{S}_i}{dt} =
-\gamma\,\mathbf{S}_i \times \mathbf{H}_{\mathrm{eff},i}
+ \frac{\alpha}{S}\,\mathbf{S}_i \times \frac{d\mathbf{S}_i}{dt},
\end{equation}
where $\gamma$ is the gyromagnetic ratio and $\alpha$ the Gilbert damping parameter. The effective field $\mathbf{H}_{\mathrm{eff},i}$ derives from exchange, anisotropy, dipolar, Zeeman, and Dzyaloshinskii--Moriya contributions.

Numerical integration schemes--semi-implicit midpoint~\cite{Daquino2005JCP}, 
Heun predictor--corrector~\cite{Sallermann2025arXiv,Leliaert2017AIPAdv}, or 
symplectic solvers~\cite{Webb2014JCP}--preserve spin magnitude and ensure stability 
over long time evolution. 
Thermal fluctuations are incorporated through stochastic fields satisfying fluctuation–dissipation relations, enabling 
simulation of temperature-dependent switching and noise-driven dynamics. In patterned systems, spatial variations in anisotropy and exchange lead to heterogeneous switching pathways. Edges, defects, and exchange bridges 
often act as nucleation sites for domain-wall formation. Spin-dynamics simulations are particularly suited for resolving domain-wall propagation, 
skyrmion motion, and current- or field-driven reversal in confined geometries.

Hybrid MC–LLG workflows are increasingly adopted: Monte Carlo sampling generates thermally equilibrated configurations, which serve as initial states for subsequent dynamic simulations. This combined strategy captures both statistical equilibrium and transient magnetization processes within a consistent Hamiltonian framework. In particular, Monte Carlo and spin-dynamics simulations provide magnetization curves, susceptibility, hysteresis loops, coercive fields, and reversal pathways, which can be compared directly with magnetometry and MOKE measurements \cite{Soldatov2024:IEEE,Lenz2021:PRApplied}.

\subsection{Multiscale and Ab Initio Parameterization}

Material-specific modeling requires reliable parameterization of exchange integrals, anisotropy constants, and Dzyaloshinskii--Moriya vectors. 
DFT provides these quantities through total-energy differences, linear-response approaches, or magnetic force theorem calculations. The resulting parameters are mapped onto effective spin Hamiltonian for atomistic simulation. Such DFT-informed spin models bridge electronic and thermodynamic scales. They enable predictive evaluation of geometry-dependent magnetic order in 2D materials and heterostructures~\cite{Aldea2025:Nanoscale}. Uncertainty in parameter extraction--arising from exchange–correlation functionals, finite supercell size, 
or Hubbard-$U$ selection—must be assessed when quantitative agreement with the experiment is sought.

At larger length scales, coarse-grained micromagnetic models approximate clusters of spins as continuous magnetization fields governed by effective material constants. 
Atomistic–micromagnetic coupling strategies allow transfer of local anisotropy or interfacial exchange information into mesoscale simulations, facilitating realistic modeling of extended patterned lattices.

\subsection{Algorithmic Advances and High-Performance Implementations}

The computational demands of patterned magnetism—particularly long-range dipolar coupling and large system sizes—have motivated substantial algorithmic development~\cite{Guan2025:arXiv}. 
GPU-accelerated implementations now enable simulation of $10^7$--$10^8$ spins with full dipolar treatment, dramatically extending accessible system sizes.

Advanced sampling methods, including parallel tempering, umbrella sampling, and transition-path techniques, 
improve exploration of metastable states and rare events such as skyrmion annihilation or domain-wall depinning. 
Sparse-field representations and FFT-based solvers further reduce computational overhead in large patterned arrays.
Machine-learning-assisted approaches are increasingly integrated with physics-based simulations. 
Surrogate models trained on Monte Carlo or spin-dynamics data can interpolate phase boundaries or
switching fields across geometric design spaces, accelerating parameter exploration while retaining physical interpretability~\cite{Kabiraj2022:Patterns, Du2025:Nanomat}. Recent solver developments, including extensible GPU-accelerated micromagnetic platforms, further expand the practical scope of large-scale magnetic simulations \cite{Moreels2026Nature}. \textit{Spin-informed graph neural networks} are recently emerging as data-efficient tools for learning magnetic ordering and accelerating materials screening \cite{Xu2025PNAS}.
Together, these computational frameworks provide a coherent toolkit for connecting microscopic Hamiltonians to thermodynamic phase diagrams and dynamic magnetization processes in patterned two-dimensional systems.

\begin{table*}[th]
\centering
\caption{Computational frameworks employed in the study of patterned two-dimensional magnetic systems. Representative references are provided for each methodological class.}
\label{tab:computational_methods}
\begin{tabular}{p{2.5cm} |p{2.75cm} |p{3.25cm} |p{3.75cm}| p{3.25cm} |p{1.5cm}}
\hline
\textbf{Framework} & \textbf{Physical Regime} & \textbf{Primary Observables} & \textbf{Principal Advantages} & \textbf{Primary Limitation} & \textbf{ Notable Ref.} \\
\hline

\textbf{Metropolis / cluster Monte Carlo}
&
(a) Canonical equilibrium sampling of spin Hamiltonians;\newline (b) Continuous thermal transitions;\newline (c) Finite-size scaling analysis
&
(a) $M(T)$, $\chi(T)$, $C(T)$, Binder cumulants, \newline (b) Structure factors, \newline (c) Correlation functions, and \newline (d) Order-parameter histograms
&
(a) Accurate evaluation of canonical ensemble averages;\newline (b) Direct access to critical exponents, pseudo-critical temperatures, and scaling collapse;\newline (c) Cluster updates mitigate critical slowing down near continuous transitions
&
(a) Local updates become inefficient near criticality;\newline (b) long-range dipolar interactions substantially increase the computational cost
&
\cite{LandauBinder2014:CUP,Wolff1989:PRL,Ferrenberg1989PRL}
\\
\hline

\textbf{Wang--Landau / multicanonical Monte Carlo}
&
(a) Generalized-ensemble sampling of rugged free-energy landscapes; \newline (b) rare-event thermodynamics; \newline (c) metastable-state exploration
&
(a) Density of states $g(E)$, \newline (b) Microcanonical entropy, free-energy barriers, \newline (c) Compensation temperatures, and \newline (d) re-entrant transition points
&
(a) Flat-histogram sampling enables uniform exploration of low-probability states;\newline (b) direct reconstruction of thermodynamic functions over broad energy windows; \newline (c) efficient for first-order-like transitions, phase coexistence, and metastability
&
(a) Convergence may be slow in large frustrated or heterogeneous systems;\newline (b) finite-size interpretation is nontrivial in patterned geometries
&
\cite{Moreno2022CompPhysCommun}\\
\hline

\textbf{Spin-dynamics/ Stochastic Landau--Lifshitz--Gilbert (LLG) dynamics} & (a) Nonequilibrium magnetization switching, \newline (b) hysteresis, \newline (c) LLG equation with thermal noise & (a) Magnetization reversal, \newline (b) hysteresis loops,\newline (c) domain-wall motion,\newline (d) skyrmion trajectories & (a) Real-time evolution including damping and thermal fluctuations; \newline (b) switching times; \newline (c) reversal pathways & (a) Expensive; \newline (b) timestep sensitive; \newline (c) not an equilibrium sampler & \cite{Lakshmanan2011:PTRSA,Muller2019:PRB} \\
\hline

\textbf{Micromagnetic simulations} & Continuum magnetization with effective-field LLG on mesoscale magnetic textures & (a) Domain configurations,\newline (b) vortex, and \newline (c) skyrmion stability & (a) Large spatial-scale modeling with effective-field description, and \newline (b) \textit{device-scale modeling} & (a) Loses atomistic resolution; \newline (b) parameter-transfer uncertainty & \cite{Vansteen2014AIPAdv} \\
\hline

\textbf{First-principles parametrization (DFT-based)}
&
Electronic-structure-driven magnetic couplings and parameter transfer
&
(a) $J_{ij}$ \newline
(b) $K$ \newline
(c) $\mathbf{D}_{ij}$ \newline
(d) Total-energy, linear-response, and magnetic force-theorem extraction
&
(a) Material-specific inputs for atomistic and continuum spin models \newline
(b) Chemically and interfacially resolved exchange, anisotropy, and DMI \newline
(c) Multiscale parameter transfer from electrons to spins
&
(a) Sensitive to XC functional, Hubbard-$U$, and supercell size \newline
(b) Costly for low-symmetry or large cells \newline
(c) Coarse-graining may introduce systematic error
&
\cite{Liechtenstein1987JMMM,Gibertini2019:Nature}
\\
\hline

\textbf{Machine-learning surrogate models}
&
High-dimensional phase mapping and inverse design
&
(a) $T_C$ \newline
(b) Coercivity \newline
(c) Switching fields \newline
(d) Geometry--property correlations
&
(a) Rapid exploration of large parameter spaces \newline
(b) Interpolation of phase boundaries and response surfaces \newline
(c) Active learning for targeted magnetic states \newline
(d) Uncertainty quantification when calibrated
&
(a) Data-hungry training \newline
(b) Weak extrapolation beyond the training manifold \newline
(c) Model interpretability depends on descriptors
&
\cite{Zhang2020:PRResearch, Nonaka2026arXiv}
\\
\hline
\end{tabular}
\end{table*}

Table \ref{tab:computational_methods} summarizes the principal computational frameworks employed in patterned
2D magnetism, highlighting their physical scope, typical observables, and representative applications. 
The various approaches play complementary roles in describing geometry-controlled spin phenomena across different length and 
time scales. Monte Carlo methods are particularly well suited for equilibrium thermodynamics, critical behavior, and phase-
diagram construction, whereas stochastic Landau–Lifshitz–Gilbert spin dynamics captures magnetization reversal, domain-wall 
motion, and skyrmion dynamics. Micromagnetic simulations extend modeling to experimentally relevant device dimensions, albeit at 
the expense of atomistic resolution, while first-principles calculations provide material-specific exchange and anisotropy 
parameters but remain computationally demanding. Consequently, modern investigations increasingly employ multiscale workflows in 
which first-principles calculations parameterize atomistic spin models that subsequently inform mesoscale micromagnetic 
simulations.

\section{Thermodynamics and Phase Behavior}
\label{sec:thermo_scaling}

The thermodynamic behavior of patterned two-dimensional magnetic systems reflects the combined effects of reduced dimensionality, geometric 
confinement, and spatially modulated interactions. These systems do not approach the thermodynamic limit in a conventional manner. Computational analysis is therefore essential for identifying effective transition temperatures, scaling behavior, and stability regimes~\cite{LandauBinder2014:CUP}. 
In patterned geometries, phase behavior becomes explicitly geometry dependent, with feature size, connectivity, 
and anisotropy contrasts acting as additional control parameters.

\subsection{Scaling and Universality}

FSS provides the principal framework for analyzing phase transitions in confined systems. Near a continuous transition, thermodynamic observables scale as
\begin{eqnarray}
M &\sim& L^{-\beta/\nu} f_M(tL^{1/\nu}), \\
\chi &\sim& L^{\gamma/\nu} f_\chi(tL^{1/\nu}), \\
C &\sim& L^{\alpha/\nu} f_C(tL^{1/\nu}),
\end{eqnarray}
where $t=(T-T_c)/T_c$ and $L$ denotes a characteristic geometric length. In patterned systems, however, $L$ must reflect effective magnetic 
connectivity rather than total system size. Weakly coupled nanodot arrays, antidot lattices \cite{Zivieri2016:AIPAdv,Belim2023:LetMat}, 
or core--shell geometries often display crossover behavior between collective ordering and quasi-independent regimes.
Deviations from standard two-dimensional universality classes may arise when exchange modulation, anisotropic connectivity, or geometric frustration 
alters the effective symmetry of the Hamiltonian~\cite{Grinstein1976:PRB,Picco2006:JourStatMech}. 
In such cases, pseudo-critical temperatures and apparent exponents depend sensitively on pattern 
periodicity and interface strength, and transitions frequently broaden into crossover regions rather than sharp singularities.

\subsection{Compensation and Reentrant Behavior}
Spatially heterogeneous exchange interactions can generate nonmonotonic magnetization profiles. In core--shell or compositionally modulated systems, 
subregions characterized by different exchange strengths or spin magnitudes respond unequally to temperature~\cite{Iglesias2007:JPhysCondMat}.
This imbalance can produce a compensation temperature $T_{\mathrm{comp}}$ at which the total magnetization vanishes while local order persists.
Reentrant behavior may also occur when competing exchange paths and dipolar interactions create entropy-driven 
reorganization of spin configurations\cite{Kabbour2026AngewChem}. 
Moderate geometric modulation can therefore stabilize ordered phases within intermediate temperature windows, producing phase diagrams qualitatively 
distinct from uniform films\cite{Streubel2016:JAPD}.

\subsection{Dynamic and Caloric Responses}

Confinement also modifies nonequilibrium thermodynamic response. Under cyclic or oscillatory magnetic fields, patterned systems often exhibit 
broadened hysteresis and geometry-dependent coercivity distributions due to edge-dominated switching pathways. Dynamic order parameters defined over 
field cycles distinguish ordered and disordered nonequilibrium regimes, with characteristic frequencies shifted by anisotropy contrasts and inter-
element coupling.

Thermodynamic response functions derived from fluctuation relations,
\begin{equation}
C = \frac{\langle E^2\rangle - \langle E\rangle^2}{k_B T^2},
\end{equation}
reveal broadened specific-heat maxima and distributed transition temperatures\cite{Binder1976PRB}. Reduced correlation volume and metastable domain 
configurations produce entropy plateaus and multi-peak heat-capacity profiles. These signatures provide quantitative measures of geometry-controlled energy barriers and thermal stability.

\subsection{Phase-Diagram Construction in Patterned Geometries}
\label{sec:phase_diagram}

A central objective of computational thermodynamics in patterned magnets is the construction of multidimensional phase diagrams spanning temperature, 
external field, anisotropy, exchange contrast, and geometric parameters~\cite{Ortiz-Ambriz2019RMP}. In contrast to homogeneous lattices, geometry 
itself functions as an effective thermodynamic variable: pattern periodicity, aspect ratio, inter-element spacing, interface coupling strength, and 
exchange modulation introduce additional axes in parameter space. Numerical studies of nanodot arrays and antidot lattices demonstrate that small 
variations in feature size or connectivity can shift ordering temperatures and alter the topology of phase 
boundaries~\cite{Wang2006Nanotechnology, Ding2011JApplPhys, Zivieri2016:AIPAdv, Sudson2021nanomaterials}. 
In confined skyrmionic systems, the stability window of chiral textures depends sensitively on disk radius and edge anisotropy, 
leading to size-dependent phase transitions between single-domain, vortex, and skyrmion 
states~\cite{Zhang2016PRB,Potkina2023PRB}.

Reliable phase mapping requires systematic averaging over disorder and finite-size effects. 
Binder-cumulant crossings, susceptibility maxima, and 
histogram reweighting analysis provide complementary indicators of transition 
boundaries~\cite{LandauBinder2014:CUP,Ferrenberg1989PRL}. For spatially 
inhomogeneous systems, phase boundaries typically broaden into finite crossover regions rather than 
sharp critical lines, particularly when exchange or anisotropy vary across subregions. 
Large-scale atomistic simulations have shown that edge roughness and interfacial exchange fluctuations generate 
distributed local transition temperatures, effectively 
renormalizing macroscopic phase behavior~\cite{Chureemart2011PRB, Evans2014JPCM}. Consequently, 
extrapolation to effective critical temperatures must account for 
connectivity-dependent magnetic length scales rather than nominal lattice dimensions.

When material-specific parameters derived from electronic-structure calculations are 
incorporated into spin Hamiltonians, phase diagrams acquire 
predictive capability. First-principles extraction of exchange interactions and 
Dzyaloshinskii--Moriya vectors enables quantitative mapping of 
stability regions in van der Waals magnets and patterned heterostructures~\cite{Mankovsky2022:ElecStr}. 
Core--shell configurations 
exhibit distinct compensation and ferrimagnetic regimes governed by interfacial exchange ratios~\cite{Iglesias2007:JPhysCondMat}, 
while micromagnetic studies of nanodisks reveal vortex-core and multi-domain stability regions controlled 
by aspect ratio and perpendicular anisotropy~\cite{Ha2003PRB, Gaididei2010PRB}. 
In patterned multilayers with interfacial DMI, confinement can stabilize isolated 
skyrmions within well-defined temperature and field windows, as confirmed by combined simulation and 
experiment~\cite{Jiang2015Science}. 

These examples illustrate that phase-diagram construction in patterned geometries is not merely descriptive but structurally integrative: geometry, 
material-specific interactions, 
and finite-size effects collectively define the accessible magnetic states. Thermodynamic mapping therefore provides 
a quantitative framework for identifying robust stability regions and for guiding controlled transitions between 
functional states under equilibrium or driven conditions.

\section{Representative Patterned Systems: Computational Insights}
\label{sec:case_studies}

The computational frameworks discussed above acquire concrete physical relevance when applied to experimentally realizable patterned 
systems. Nanodot arrays, artificial spin-ice lattices, core--shell heterostructures, and patterned van der Waals magnets provide 
distinct platforms in which geometry and interaction engineering reshape magnetic phase stability. These systems illustrate how 
thermodynamic phase mapping, finite-size scaling, and multiscale parameterization converge in practice.

\subsection{Nanodot and Antidot Arrays}

Nanodot and antidot arrays constitute prototypical patterned magnets in which confinement and dipolar coupling govern collective behavior. Circular nanodisks support vortex states that minimize magnetostatic energy, with stability regions determined by disk radius, thickness, and perpendicular anisotropy \cite{Shinjo2000Science,Guslienko2001PRB}. Micromagnetic and atomistic simulations reveal size-dependent transitions between single-domain, vortex, and multidomain configurations, producing geometry-modulated phase diagrams in the temperature--field plane \cite{Cowburn1999PRL}.

Antidot lattices, formed by periodic perforation of a continuous film, introduce anisotropic connectivity and modified demagnetization 
fields. Computational studies demonstrate that antidot diameter and spacing tune coercivity and shift effective Curie temperatures 
through reduced coordination and enhanced boundary pinning \cite{Zivieri2016:AIPAdv}. Large-scale Monte Carlo 
simulations further show that positional disorder broadens phase transitions and generates distributed blocking temperatures. These 
findings establish nanodot and antidot systems as model platforms for investigating confinement-driven magnetic scaling and crossover 
behavior.

\subsection{Artificial Spin-Ice Lattices}

Artificial spin-ice arrays extend geometric control to frustrated magnetism. Lithographically defined nanoislands arranged in square or 
kagome geometries emulate the ice rules of pyrochlore systems \cite{Wang2006Nature}. In these systems dipolar interactions dominate the 
effective Hamiltonian, and Monte Carlo simulations reproduce emergent monopole-like excitations and Dirac-string correlations \cite{Moller2006PRL,Nisoli2013RMP}.
Recent computational studies have mapped phase boundaries separating ordered, partially ordered, and disordered regimes as a function of lattice spacing, island aspect ratio, and thermal activation \cite{Farhan2013PRL,Ortiz-Ambriz2016}. Finite-size analyses indicate crossover behavior between collective ordering and superparamagnetic island response, depending on inter-element coupling strength. Artificial spin ice therefore serves as a benchmark system for testing large-scale dipolar Monte Carlo methods and rare-event sampling approaches in geometrically frustrated lattices.

\subsection{Core--Shell and Exchange-Modulated Heterostructures}
Core--shell and laterally modulated heterostructures introduce spatially varying exchange and anisotropy profiles within a single 
patterned unit. Monte Carlo simulations of core--shell nanoparticles reveal compensation temperatures and reentrant behavior controlled 
by interfacial exchange ratios \cite{Iglesias2005PRB,Kachkachi2005}. In two-dimensional patterned films, exchange modulation between 
regions of differing thickness or composition generates multiple magnetic transitions and broadened susceptibility peaks.
Micromagnetic simulations further demonstrate that interfacial DMI gradients stabilize localized chiral 
textures within confined regions \cite{Jiang2015Science}. When exchange parameters are derived from first-principles calculations and 
incorporated into atomistic spin models, computed phase diagrams capture material-specific stability windows for ferrimagnetic, vortex, 
or skyrmionic states \cite{Matthias2024PhysRevResearch}. These systems highlight the predictive capacity of combining 
electronic-structure inputs with thermodynamic spin modeling.

\subsection{Patterned van der Waals Magnets}
The discovery of intrinsic two-dimensional magnets such as CrI$_3$, Cr$_2$Ge$_2$Te$_6$, and Fe$_3$GeTe$_2$ has extended patterned 
magnetism to the atomic limit \cite{Huang2017Nature,Gong2017Nature}. Nanostructuring of these materials through lithography, strain 
engineering, or electrostatic gating modifies exchange anisotropy and stabilizes novel spin textures. Atomistic spin simulations 
predict thickness- and geometry-modulated Curie temperatures consistent with experimental magneto-optical Kerr measurements \cite{Soriano2020NanoLett}.
In patterned Fe$_3$GeTe$_2$ nanoflakes, confinement enhances perpendicular anisotropy and shifts vortex--skyrmion phase boundaries 
under applied fields \cite{Zhang2025MatToday}. Earlier work established that machine learning can identify magnetic ordering in 2D materials and provide chemically interpretable descriptors for magnetic screening \cite{Acosta2022ACSAMI}. Combined DFT and spin-dynamics studies reveal that edge termination and symmetry breaking strongly influence DMI and magnetocrystalline anisotropy, reshaping thermodynamic stability regions~\cite{Jaeschke2021PRB}. Recent experiments on twisted two-dimensional antiferromagnets further underscore the richness of geometry-controlled magnetic states in atomically thin systems \cite{Wong2026NatNano}.
Patterned van der Waals magnets, therefore, represent a regime in which reduced dimensionality, 
electronic structure, and geometric confinement are intrinsically coupled.
Across these representative systems, computational modeling establishes a consistent picture: geometry, connectivity, and interfacial 
coupling enter the magnetic Hamiltonian on equal footing with intrinsic exchange interactions. Phase-diagram construction thus emerges as a unifying framework linking nanoscale structure to stabilized magnetic states and functional switching behavior.\\

Domain patterns, skyrmions, and other nontrivial textures predicted by micromagnetic and atomistic models can be visualized using Lorentz TEM, MOKE microscopy, magnetic force microscopy, and spin-polarized scanning probe techniques \cite{Nguyen2022:PRApplied,Wortmann2001:PRL}.

\section{Emerging Computational Paradigms and Future Directions}
\label{sec:comp_trends}

The computational study of patterned two-dimensional magnetic nanostructures is evolving from equilibrium modeling toward integrated, predictive simulation frameworks. Increasing geometric complexity, material specificity, and device relevance demand more advanced approaches. These approaches must incorporate stochastic dynamics, multiphysics coupling, quantum corrections, and scalable data-centric workflows. The following 
directions outline the principal methodological frontiers shaping the next phase of computational patterned magnetism.

\subsection{Nonequilibrium and Stochastic Magnetism}

Patterned nanomagnets typically operate under thermally fluctuating and externally driven conditions. The stochastic Landau--Lifshitz--
Gilbert (sLLG) formalism, incorporating Langevin noise consistent with the fluctuation–dissipation theorem, provides a microscopic 
description of switching-time distributions, retention statistics, and telegraph noise \cite{Muller2019:PRB}, even large-scale simulations now can be performed in limited time and on inexpensive hardware\cite{Vansteen2014AIPAdv}. 
Rare events such as domain-wall depinning\cite{Keck2025}, vortex annihilation\cite{Arekapudi2021PRB}, or skyrmion collapse\cite{Derras-Chouk2022PRB} often dominate functional stability. 
Geodesic nudged elastic band (GNEB) methods combined with harmonic transition-state theory and 
kinetic Monte Carlo (KMC) now enable quantitative 
lifetime prediction across temperature ranges \cite{GPUMonteCarlo2021}. Transition-path sampling and forward-flux techniques further 
resolve complex switching pathways beyond single-barrier approximations. 
These developments extend equilibrium phase diagrams into kinetic phase space, where metastable lifetimes and entropy production define 
stability metrics relevant for device operation.

\subsection{Multiphysics and Hybrid Coupling}
Patterned magnets are inherently coupled to mechanical, electrical, and optical degrees of freedom. Magnetoelastic and magnetoelectric 
interactions introduce additional control axes—strain and voltage—into thermodynamic phase space~\cite{Teh2023Nanomaterials}. 
Finite-element micromagnetics coupled 
to elasticity solvers reproduce strain-stabilized skyrmions and voltage-tunable anisotropy \cite{Hu2018npjComputMat,Misba2025ApplMatInt}. DFT-informed spin dynamics predicts field-driven exchange modulation in gated heterostructures~\cite{Mankovsky2022}. Recent solver developments, including extensible GPU-accelerated micromagnetic platforms, further expand the practical scope of large-scale magnetic simulations \cite{Moreels2026Nature}. Hybrid magnon–photon and magnon–phonon systems further require coupling between spin dynamics and electromagnetic solvers~\cite{Berk2019NatureCommun,Zhang2025npjSpin}. 
Effective Hamiltonian approaches and time-domain simulations capture mode hybridization and 
strong-coupling regimes, linking patterned magnetism 
to quantum information and microwave photonics. Multiphysics integration therefore transforms patterned magnets into coupled-field 
systems whose stability must be analyzed within extended parameter spaces.

\subsection{Quantum and Beyond-Classical Modeling}
As patterned structures approach atomic thickness and strong spin–orbit coupling regimes, purely classical spin descriptions become insufficient. 
Quantum Monte Carlo, tensor-network methods, and exact diagonalization are increasingly applied to anisotropic Heisenberg and Kitaev-
type models relevant to two-dimensional magnets~\cite{Hait2025CUDA,Zou2025PRB}. 
These approaches capture quantum fluctuations, entanglement structure, and 
low-temperature ordering phenomena beyond classical approximations.
In Moir\'e-patterned or twisted magnetic bilayers, emergent superlattice periodicities introduce additional quantum interference and 
magnon band reconstruction~\cite{Wang2024NatCommun}.
Bridging quantum spin models with classical thermodynamic simulations through renormalized effective 
Hamiltonians remains an open problem, particularly for finite-temperature patterned systems. Developing consistent quantum-to-classical 
coarse-graining strategies is therefore a central theoretical challenge.

\subsection{Scalable, Data-Centric, and Predictive Simulation Ecosystems}
Scalability remains a defining constraint for realistic patterned systems containing long-range dipolar and Dzyaloshinskii--Moriya 
interactions. GPU-accelerated Monte Carlo and spin-dynamics codes now scale to millions of spins, while hierarchical coarse-graining 
bridges atomistic and micromagnetic regimes \cite{Chen2024JCompSci, Muller2019:PRB}. 
Parallel to algorithmic advances, data-centric integration has been becoming essential~\cite{Rhone2020SciRep}. 
Interpretable machine-learning surrogates and active-learning strategies accelerate phase mapping across 
high-dimensional geometric parameter spaces~\cite{Acosta2022ACSAMI}. 
Bayesian uncertainty quantification provides confidence intervals for predicted transition 
temperatures and coercive fields, addressing fabrication-induced variability~\cite{Hu2025PRM}. 
Emerging digital-twin concepts aim to integrate electronic-structure calculations, thermodynamic spin simulations,
and experimental imaging data into adaptive, continuously updated models of patterned devices\cite{Bolyachkin2024CompMat}. Machine-learning-assisted screening has also been applied to intrinsically spin-ordered 2D nanomagnets, demonstrating the utility of interpretable models for rapid property estimation \cite{Kar2024ACSAMI}. So we realize such ecosystems will require standardized data formats, reproducible workflows, and interoperability between atomistic and continuum solvers.
Collectively, these developments indicate a transition from equilibrium-focused modeling toward integrated, uncertainty-aware, and 
multiscale simulation platforms. The extent to which computational patterned magnetism achieves predictive design capability will 
depend on scalable algorithms, consistent parameter transfer across scales, and rigorous validation against experiment.\\
For clarity, Table~\ref{tab:emerging_paradigms} compiles the emerging computational paradigms outlined in this section, identifying 
their scientific objectives, methodological challenges, and representative solution strategies. The overview underscores the ongoing 
transition from descriptive simulation toward scalable and uncertainty-aware predictive modeling.
\begin{table*}[ht]
\centering
\caption{Emerging computational paradigms in patterned two-dimensional magnetism and their associated challenges.}
\label{tab:emerging_paradigms}
\begin{tabular}{p{3cm}| p{3.2cm}| p{3.2cm}| p{3.2cm}| p{2cm}}
\hline
\textbf{Paradigm} & \textbf{Scientific Objective} & \textbf{Primary Challenge} & \textbf{Computational Strategy} & \textbf{Representative References} \\
\hline

Nonequilibrium stochastic modeling & Thermally activated switching and noise characterization & Rare-event timescales beyond direct simulation & sLLG integration, transition-path sampling, forward-flux methods & \cite{Muller2019:PRB,Guan2025:arXiv} \\

Multiphysics coupling & Strain- and voltage-controlled magnetism & Coupling spin, elastic, and electronic degrees of freedom & Finite-element micromagnetics with DFT-informed parameters & \cite{Smidstru2020JPhysCondMatt,Salehi2025SciRep} \\

Quantum spin simulations & Quantum fluctuations and entanglement effects & Sign problem and exponential Hilbert-space growth & Quantum Monte Carlo, tensor networks, Lindblad dynamics & \cite{Kabiraj2022:Patterns, Grubisic-cabo20252DMat} \\

Data-centric workflows & Accelerated phase mapping and inverse design & High-dimensional parameter exploration & Active learning, Gaussian processes, physics-informed neural networks & \cite{Kovacs2022JMMM,Gong2023IEEE, McClenny2023JCompPhys, Cao2025ChemMat} \\

Exascale and GPU computing & Large-scale dipolar and long-range interactions & Memory bandwidth and communication bottlenecks & Hybrid MPI+CUDA frameworks and optimized FFT solvers & \cite{Vansteen2014AIPAdv,Lepadatu2023JApplPhys,Nonaka2026arXiv} \\

\hline
\end{tabular}
\end{table*}

\section{Conclusion}
\label{sec:conclusion}

Patterned two-dimensional magnetic systems constitute a class of architected spin materials in which geometry, exchange interactions, anisotropy, and topology operate on comparable energy scales. Spatial modulation of magnetic lattices transforms otherwise uniform films into confined, heterogeneous systems where geometry serves as an intrinsic control variable for their thermodynamic and dynamical behavior. 

Computational approaches—ranging from equilibrium Monte Carlo sampling to stochastic spin dynamics and multiscale parameterization 
based on first-principles calculations—provide the quantitative framework required to resolve this complexity. These methods enable 
construction of geometry-controlled phase diagrams, identification of compensation and reentrant phenomena, and analysis of metastable 
switching pathways in nanostructured architectures. Thus, representative systems such as nanodot arrays, artificial spin-ice lattices, core-shell heterostructures, and patterned van der Waals magnets provide distinct platforms. In these systems, geometry and interaction engineering reshape magnetic phase stability.

A central insight emerging from this study is that geometry functions as an effective thermodynamic variable. Pattern 
periodicity, aspect ratio, exchange contrast, and boundary conditions modify correlation lengths, broaden transition regimes, and 
define stability windows for vortices, skyrmions, and compensated states. Phase-diagram construction therefore serves as a systematic 
bridge between microscopic Hamiltonians and experimentally realizable magnetic configurations.

Despite substantial progress, predictive accuracy remains limited by challenges associated with disorder, long-range dipolar interactions, and multiscale coupling across electronic, atomistic, and continuum levels. Extending equilibrium thermodynamics to nonequilibrium, multiphysics, and stochastic regimes continues to demand algorithmic refinement and careful uncertainty quantification. The development of scalable, reproducible, and interoperable simulation workflows will be decisive for translating computational insight into robust device-level design.

In essence, patterned 2D magnetism exemplifies the convergence of geometry-controlled materials engineering and computational 
statistical physics. As simulation frameworks become more integrated and experimentally validated, the capacity to identify and 
stabilize targeted magnetic states through geometry and interaction tuning is expected to strengthen. The field is moving from 
descriptive modeling toward quantitatively reliable, design-oriented computation. This progression defines the theme of this review 
article.

\begin{acknowledgments}
This work was supported by the Anusandhan National Research Foundation (formerly Science and Engineering Research Board) 
through the start-up research grant (SRG/2023/000122).
\end{acknowledgments}

\bibliography{References_rev_patterned2d.bib}

\end{document}